# Circularly polarised electroluminescence from chiral excitons in vacuum-sublimed supramolecular semiconductor thin films

Rituparno Chowdhury[1#], Marco D. Preuss[2#], Hwan-Hee Cho[1], Joshua J. P. Thompson[3], Samarpita Sen[4], Tomi Baikie[1], Pratyush Ghosh[1], Yorrick Boeije[1], Xian-Wei Chua[1,5], Kai-Wei Chang[3], Erjuan Guo[6], Joost van der Tol[2], Bart W. L. van den Bersselaar[2], Andrea Taddeucci[7,8], Nicolas Daub[2], Daphne M. Dekker[9], Scott T. Keene[1], Ghislaine Vantomme[2], Bruno Ehrler[9], Stefan C. J. Meskers[2], Akshay Rao[1], Bartomeu Monserrat[3], E. W. Meijer[2]*, Richard H. Friend[1]*

[1] Cavendish Laboratory, University of Cambridge, J. J. Thomson Avenue, Cambridge, CB3 0HE, United Kingdom.

[2] Institute for Complex Molecular Systems and Laboratory of Macromolecular and Organic Chemistry, Eindhoven University of Technology, P.O. Box 513, 5600 MB Eindhoven, The Netherlands.

[3] Department of Materials Science and Metallurgy, University of Cambridge, 27 Charles Babbage Road, Cambridge CB3 0FS, United Kingdom

[4] The Gurdon Institute and the Department of Genetics, University of Cambridge, Cambridge CB2 1QN, United Kingdom.

[5] Department of Chemical Engineering and Biotechnology, University of Cambridge, Philippa Fawcett Drive, Cambridge CB3 0AS, United Kingdom.

[6] State Key Laboratory of Materials Processing and Die and Mould Technology, School of Materials Science and Engineering, Huazhong University of Science and Technology, Wuhan, China

[7] B23 Beamline, Diamond Light Source Ltd, Didcot OX11 0DE, United Kingdom.

[8] Dipartimento di Chimica e Chimica Industriale, University of Pisa, Via Moruzzi 13, 56124 Pisa, Italy.

[9] Center for Nanophotonics, AMOLF, Science Park 104, 1098 XG Amsterdam, The Netherlands.

[#] These authors contributed equally.

*e-mail: e.w.meijer@tue.nl, rhf10@cam.ac.uk

**Abstract:**
Materials with chiral electronic structures are of great interest. We report a triazatruxene, TAT, molecular semiconductor with chiral alkyl side chains that crystallises from solution to form chirally-stacked columns with a helical pitch of 6 TATs (~2.3 nm). These crystals show strong circularly polarised, CP, green photoluminescence, with dissymmetry of 24%. Electronic structure calculations using the full crystal structure, show that this chiral stacking associates angular momentum to the valence and conduction states and thus gives rise to the observed CP luminescence. Free-standing crystals are not useful for active semiconductor devices, but we have discovered that co-sublimation of TAT as the 'guest' in a structurally mismatched 'host' enables the fabrication of thin films where the chiral crystallization is achieved in-situ by thermally-triggered nano-phase segregation of dopant and host whilst preserving the integrity of the film. This enables fabrication of bright (green) organic light-emitting diodes with unexpectedly high external quantum efficiencies of up to 16% and electroluminescence dissymmetries ≥10%. These materials and this process method offer significant application potential in spintronics, optical displays and multidimensional optoelectronics.

**Introduction**
Obtaining chiral nanostructures inside semiconductor devices is inherently complex due to the challenge of manipulating molecular arrangements with precise chirality at the nanoscale (1). Maintaining uniform chirality across large areas is a further challenge using well established device fabrication methods such as vacuum sublimation with no reports on this so far. The dynamic and non-covalent nature of supramolecular interactions enables the fabrication of chiral structures following a bottom-up approach (1–5) by transferring asymmetry from the molecular to the nanoscopic/microscopic scale (6). This has previously enabled access to chiral materials for advanced spintronic and optoelectronic applications (7). A notable example is organic light-emitting diodes (OLEDs) where the emergence of circularly polarised electroluminescence (CP-EL) from chiral semiconductors gives rise to broad applications in 3D displays (8), telecommunication (9) and high-contrast microscopy (10).

Since the first reports of a circularly polarized organic light emitting diode (CP-OLED) (11, 12), significant efforts (13–16) have been devoted towards improvements in device efficiency and the degree of dissymmetry in CP-EL ($g_{EL}$). It is possible to obtain significant $g_{EL}$ from inorganic quantum-well-based device architectures has been shown under high magnetic fields or at low temperatures. For organic semiconductors, strong dissymmetry in the electroluminescence, EL, close to the theoretical maximum of $g_{EL}$ = ±2, was mostly realized for materials forming chiral, cholesteric supramolecular aggregates (14). However, the overall performance of such materials suffers from dark triplet states that limit the internal quantum efficiency (IQE). Chiral heavy-metal complexes and thermally activated delayed fluorescence (TADF) based CP-OLEDs (16, 17) have been shown to overcome this triplet problem with IQE's close to unity, but suffer from very low dissymmetries with $g_{EL}$ ~ $10^{-3}$. This observed trade-off between device performance and EL dissymmetry limits the application of CP-OLEDs to date.

In all of the described materials, CP-EL can originate locally from the emission of circularly polarised photons molecular sites, or non-locally from the selective circularly-polarised scattering of photons in a cholesteric medium. Both strategies can achieve large dissymmetries ($g_{EL}$ ~ 1) but do not yield high device efficiencies (EQE < 1%). In this report we combine high dissymmetries in EL with high device performance. To this end, we exploit thermally triggered phase segregation in vacuum sublimed host-dopant films to fabricate chiral supramolecular nanostructures. To our surprise, the supramolecular assembly enables highly dissymmetric emission ($g_{EL}$ ~ $10^{-1}$) while providing access to efficient ($EQE_{max}$ ~ 16%) and bright ($L_{max}$ > 5x$10^4$ cd m$^{-2}$) devices.

**Supramolecular assembly and structure**
The molecules synthesized, shown in Figure 1(a), share the triazatruxene (TAT) core with different N-substituents: *S*-3,7-dimethyloctyl (*S*-TAT), *n*-octyl (*n*-TAT) and phenyl (*phen*-TAT). Out of these molecules only *S*-TAT possesses intrinsic chemical chirality. The remaining 2 molecules, *n*-TAT and *phen*-TAT, serve as non-chiral and aromatic structural controls to study the emergent chiral supramolecular behaviour of *S*-TAT which is described later. Free standing single crystals were obtained by slow evaporation from toluene:hexane solution. The crystal structure of *S*-TAT, shown in Figure 1(b), exhibits a non-centrosymmetric packing with a primitive unit cell conforming to the chiral, *P*6$_1$22 space group. When viewed along the *b*-axis we see that the *S*-TAT dimerises and organize helically through co-facial π-stacking and exhibit a spacing of 0.38 nm in the dimer and 0.39 nm in between dimers along the *a*-axis. Viewing along the *c*-axis reveals the 6-fold rotation axis of the stacking and also shows that the molecules are face-to-face stacked without any slippage. From this structure we can determine the helical pitch of the supramolecular assembly to be 2.3 nm and an intermolecular rotation angle of 40° within the dimer and 120° in-between dimers. The crystal structure of other TAT derivatives and S-TAT polymorphs are discussed in the Supporting Information. We also fabricated vacuum co-sublimed thin films of TAT with molecular semiconductor hosts, including carbazole biphenyl, CBP. As presented later, *in situ* crystallisation of the TAT within the host matrix is readily achieved, which is triggered by the shape dissimilarity between the *N*-substituent of TAT and the semiconducting molecular host as illustrated schematically in Figure 1(c). These thin films are used as the emitting layers in LEDs, as discussed below.

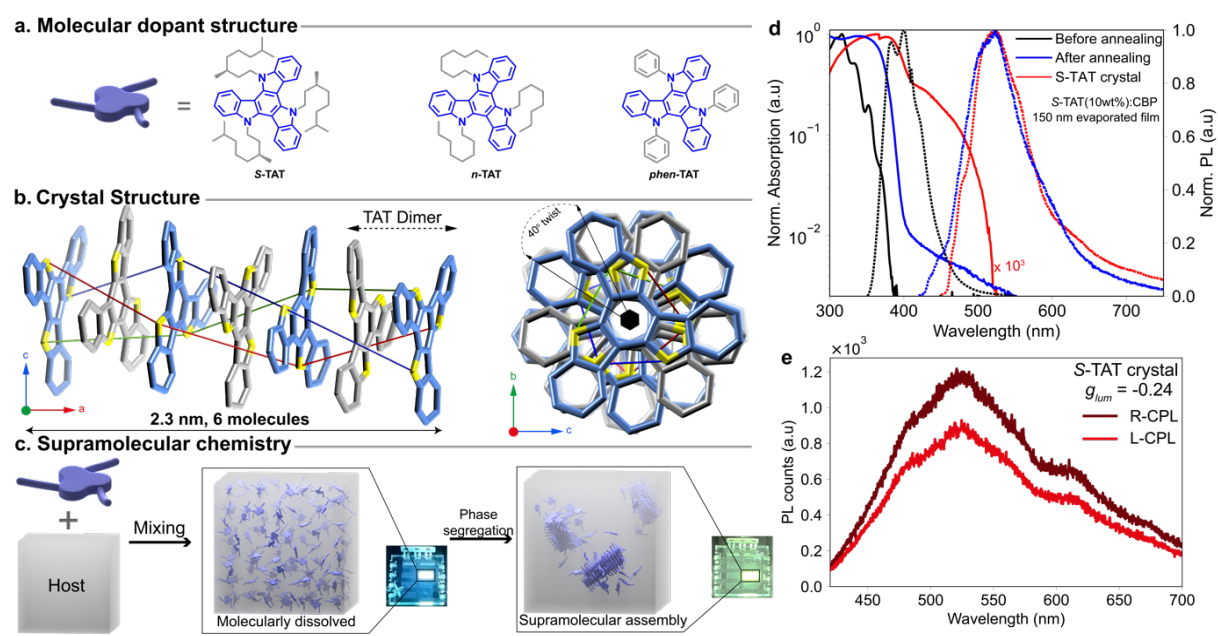

**Figure 1: Supramolecular chiral assembly of chromophores. (a)** The triazatruxene molecules synthesized and studied. *S*-3,7-dimethyloctyltriazatruxene (*S*-TAT), n-octyltriazatruxene (*n*-TAT), phenyltriazatruxene (*phen*-TAT). **(b)** The crystal structure of *S*-TAT. *(left)* Viewed along the *b*-axis of the unit cell, the components of the *S*-TAT dimer are coloured blue and grey, this is the repeating unit of the helix. There are 6 molecules in a full-turn of the helix which extends along a pitch of 2.3 nm. *(right)* View along the *a*-axis of the unit cell. **(c)** Supramolecular assembly of TAT within a molecular semiconductor host; a cartoon of the proposed arrangement and a picture of the device. **(d)** Optical absorption measured by photothermal deflection spectroscopy (solid lines) and emission (dashed lines) of the *S*-TAT pure crystal (red), the 150 nm thick vacuum sublimed thin film of 10wt% TAT in CBP before annealing (black), after annealing (blue). Further analysis of the spectra is provided in the Supporting Information. **(e)** Circularly polarised photoluminescence (CP-PL) spectrum of the pure S-TAT crystal. We observe a CP-PL dissymmetry, $g_{lum}$, of -0.24. We have used a 400 nm pulsed laser excitation at a fluence 5 µJcm$^{-2}$ throughout.

Optical absorption measured using photothermal deflection spectroscopy (PDS) and photoluminescence (PL) spectroscopy results for both single crystals and films of *S*-TAT are shown in Figure 1(d). Pristine vacuum-sublimed, amorphous thin films of *S*-TAT show that the isolated non-interacting molecule has a sharp absorption edge at 380 nm, displaying PL with vibronic features at 390 nm and 407 nm with a photoluminescence quantum efficiency (PLQE) of 14%. The *S*-TAT chiral crystal has its absorption edge red-shifted by ~0.65 eV to 510 nm and has a PLQE of 58%. As shown in Table S.1 this PLQE is maintained for the green emission in annealed vacuum-sublimed thin-films. These steady state optical properties are summarised in Table S.1.

Pressure dependent measurements of PL spectra on single crystals and thin-films, up to 350 MPa, are shown in Figure S.22-23. There is a significant reduction in PL intensity with increasing pressure (up to 16% at 350 MPa) and some broadening of the spectra, but relatively little shift in the peak PL energy. These are small changes, and we note that red-shifts are seen at higher pressures in π-stacked organic systems(18).

We perform measurements of circularly polarised photoluminescence (CP-PL) by distinguishing between the left-handed (L-CPL) and right-handed (R-CPL) components of emitted photons using a super-achromatic quarter waveplate followed by a linear polariser that serves as an analyser (see Methods) (19). The CP-PL spectrum of the *S*-TAT crystals is shown in Figure 1(e). The spectrum is centred at 540 nm which clearly shows a difference between the L-CPL and R-CPL components with $g_{lum}$ = -0.24. Plots of the intensity against the angle of the rotating quarter waveplate have been shown for all relevant systems in Figure S.9.

**Electronic Structure and chiral emission.**

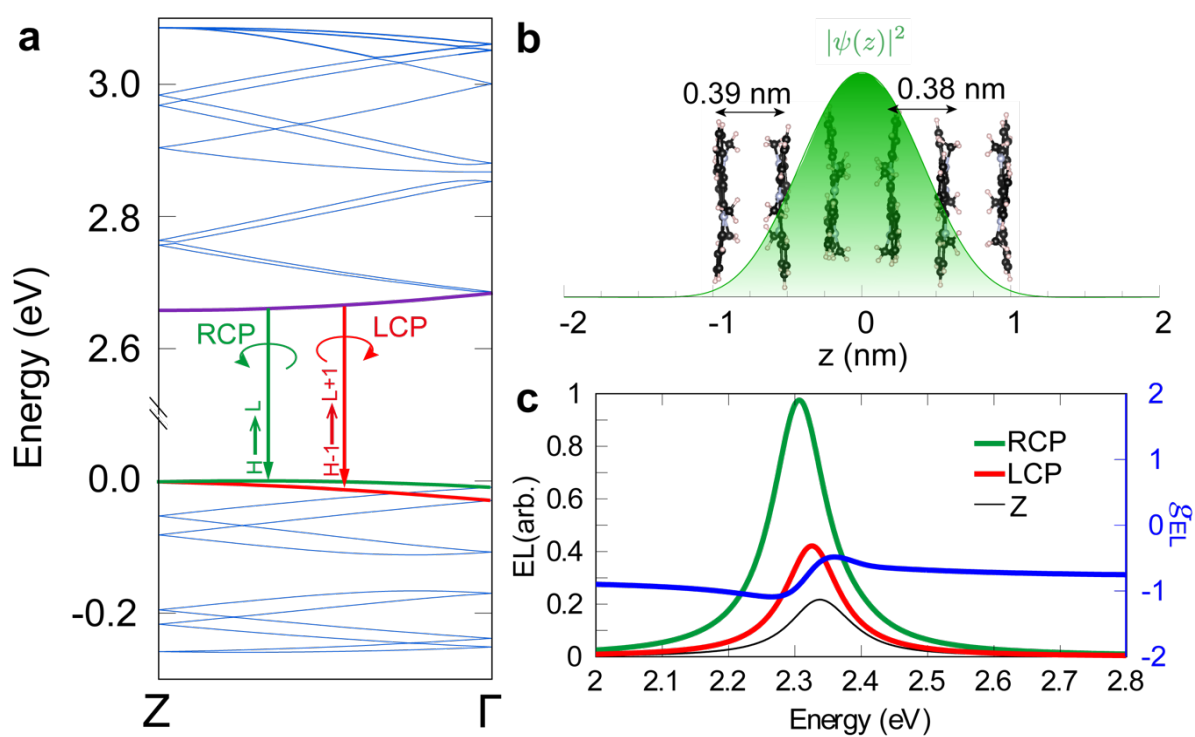

**Figure 2:** *ab-initio* **calculations on the chiral electronic structure of *S*-TAT**. **(a)** Electronic band structure of stacked-TAT (blue) with lattice spacing, *a* = 0.39 nm, thus Z = π/6*a*. The HOMO (LUMO) and HOMO-1 (LUMO+1) are highlighted in orange and red respectively. Relevant optical transitions are marked with vertical arrows with their circular polarisation indicated by the curved arrow. **(b)** Real-space excitonic wavefunction, $|\psi(\mathbf{R})|^2$, overlaid on to unit cell, showing the exciton is spread over multiple molecules. **(c)** Modelled luminescence spectrum at 300 K of right-handed (orange), left-handed (red) and Z (black) polarised light. The calculated dissymmetry factor is shown in blue.

We perform first principles electronic calculations on a single chiral stack of *S*-TAT; for computational ease we reduce the alkyl sidechains to methyl groups (for details see Methods). We find that, moving from the isolated TAT molecule to the stack, the energy of the π-π* gap reduces from 3.2 eV to 2.67 eV, a redshift of around 0.53 eV, owing to the significant π-π intermolecular stacking with relatively strong orbital overlap and strong dipole-dipole interactions. We attribute the experimentally observed red-shift from blue to green emission on crystallization to these intermolecular interactions. The intermolecular coupling gives rise to a dispersive electronic band structure as shown in Figure 2(a).

Isolated TAT molecules possess doubly degenerate HOMO (H) and LUMO (L) energy levels. In the stacked phase, shown in Figure 2(a), these energy levels are degenerate at the Z point however the intermolecular coupling lifts their degeneracy with the two bands dispersing differently. The two degenerate levels carry opposite magnetic quantum number, $m_l = \pm 1$, and couple differently to their twisted neighbouring molecule due to differences in the orbital overlap between neighbouring molecules. The two π and two π* bands are folded back within the reduced Brillouin zone to give six bands, which are gapped between the third and fourth bands due to the dimerization, opening a gap at -0.12 eV and at 2.9 eV, in the HOMO and LUMO respectively.

The two π HOMO and two π* LUMO bands would be degenerate in the absence of a chiral structure, as they are for the isolated molecule, shown in Figure S.36. However, in the presence of the chiral inter-molecular π and π* contacts, this degeneracy is lifted, to form two bands with L=1, with energy separated $m_L$ = +1 and $m_L$ = -1 bands. These disperse differently at general k values in the Brillouin zone, but converge at the band minima and maxima, as seen in Figure 2(a). We also observe an additional band at 2.88 eV, stemming from the LUMO+2 orbital. This band couples with the neighboring LUMO and LUMO+1 orbitals leading to an avoided crossing.

We can extract the optical selection rules of each pair of bands via the momentum matrix element, $\langle \psi_f | \hat{p} | \psi_i \rangle$. We find that the lowest energy transition H →L carries right-handed circular polarization (RCP) while the H-1→L+1 transition carries left-handed circular polarization (LCP), shown in Figure 2(a). These selection rules can be understood from the change in magnetic quantum number $\Delta m$, between the initial and final state. Note that the magnetic quantum number is defined modulo 3 (due to $C_3$ symmetry), and hence $\Delta m_l = \pm 2$ is equivalent to $\Delta m_l = \pm 1$. We find that for the H→L (or H-1→L+1), $\Delta m_l = +1(-1)$, corresponding to RCP (LCP). In contrast, we find that H→L+1 and H-1→L result in $\Delta m_l = 0$, corresponding to Z-polarisation.

The Wannier equation approach is used to calculate the excitonic envelope functions (20, 21), shown in Figure 2(b). We find that the singlet, $S_{1,stack}$, exciton is spread over 4 molecules with a Bohr radius ~0.5 nm that is larger than the intermolecular separation; this suggests delocalisation of the exciton along the TAT stack. Near the band gap, the weaker intermolecular coupling for the HOMO is clearly lower than for HOMO-1. As such, the exciton is composed of a hole in the HOMO that is much more spatially localised than the hole in the HOMO-1 level, leading to distinct excitonic resonances for the HOMO→LUMO and HOMO-1→LUMO+1 excitons.

Then the Elliot formula (20, 21) (see Methods) is used to calculate PL spectra of the TAT stack, shown in Figure 2(c). Assuming the exciton population is thermalised, we find that the R-CPL is larger than the L-CPL and Z-polarised PL. The calculated $g_{lum}$ is shown in Figure 2(c) showing clearly an expected CP-PL. We attribute the CP-PL to the larger chiral stacking modulated intermolecular-coupling of the HOMO-1. In a single molecule there is no intermolecular coupling, hence the L-CPL and R-CPL transitions happen with equal probability, leading to negligible CPL. We explore other structural arrangements of the TAT molecules (See Supporting Information Section 4.2) which may be present due to stacking faults, this includes a non-dimerised structure (Figure S.35-36). We find that these alternate structures give neither CPL nor do they exhibit bright optical transitions.

Our model captures the origin of the large red-shift and large circular-polarisation at the supramolecular level: The intermolecular π-π stacking extends the exciton across multiple molecules thus lowering its energy, while the extension across multiple molecules can generate a structurally imposed lifting of the exciton degeneracy resulting in defined angular momentum selection rules that allows for handed-photons to be emitted from the chiral supramolecular assembly.

**Vacuum Sublimation and fabrication of chiral thin films.**

For practical purposes, single crystals are not feasible. We perform vacuum co-sublimation to fabricate thin films of 10wt% of the *S*-TAT molecule was performed alongside a large selection of *p*-type and *n*-type organic semiconducting hosts (see Table S.2). We focus our description on the results obtained from thin films of 10wt% *S*-TAT doped in 4,4'-Bis(N-carbazolyl)-1,1'-biphenyl (CBP) and 3',5'-Di(carbazol-9-yl)-[1,1'-biphenyl]-3,5-dicarbonitrile (DCzDCN). CBP is a hole-transporting semiconductor(22) while DCzDCN is an ambipolar semiconductor (23, 24). Crystallisation in all systems can be triggered through annealing above the $T_m$ of *S*-TAT around 353 K shown in Figure S.33, this leads to the formation of spherulitic domains in the thin-films whose development can be tracked through polarised optical microscopy, shown in Figure S.26. Small angle X-Ray scattering (SAXS) confirmed that thermally triggered phase segregated structures were formed in the bulk, shown in Figure S.24. Atomic force microscopy (AFM) and scanning electron microscopy (SEM) studies show that these spherulites are built up from 3-dimensional interpenetrated long crystalline structures that run parallel to the substrate (see Figures S.26-28).

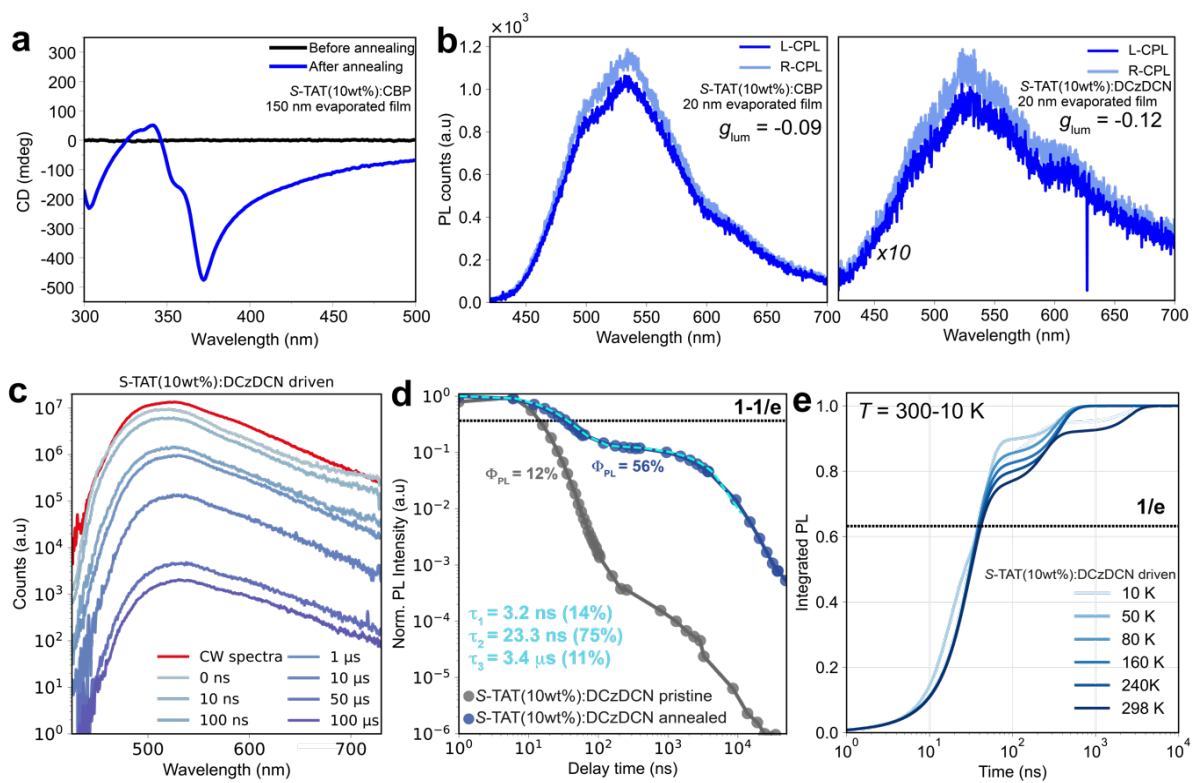

**Figure 3: Photophysics of the vacuum sublimed S-TAT doped thin-films. (a)** Circular dichroism, CD, spectra obtained by averaging the Muller matrix CD component measured on vacuum-sublimed *S*-

TAT(10%):CBP thin films before annealing (black) and after annealing (blue). For the full Muller matrix imaging see Figure S.27. **(b)** CP-PL spectrum on annealed, vacuum sublimed 20 nm thick films of (left) *S*-TAT(10%):CBP and (right) *S*-TAT(10%):DCzDCN with a $g_{lum}$ = -0.09 and $g_{lum}$ = -0.12, respectively. A fluence of 12 µJcm$^{-2}$ for the CBP films and 2.0 µJcm$^{-2}$ for the DCzDCN films is used, in both cases a 400 nm pulsed laser excitation was used which lies 0.15 eV below the first excited state of amorphous and molecular *S*-TAT. **(c)** Transient photoluminescence (TRPL) spectrum measured on 20 nm thick *S*-TAT(10%):DCzDCN films. The PL shifts from 520 nm to 530 nm within 10 ns. The emission persists beyond 100 µs at room temperature. The continuous wave (cw) PL spectrum is shown in red. **(d)** TRPL kinetic traces of the net emission observed from 20 nm thick vacuum sublimed *S*-TAT(10%):DCzDCN films before annealing (light-blue, $l_{PL}$ = 400-450 nm) and after annealing (dark-blue, $l_{PL}$ = 420-700 nm). The triexponential decay model fitting curve (cyan dashed-lines) to the annealed film PL kinetics is overlaid on the raw kinetics and the lifetimes are quoted. **(e)** Integrated Photoluminescence time-traces plotted as a function of temperature for 20 nm thick vacuum sublimed *S*-TAT(10%):DCzDCN films in the 10 K – 298 K range. A 400 nm pulsed laser excitation for annealed films at a fluence 2 µJcm$^{-2}$ in measurements (c)-(e) is used. For the pristine films in (d) a 350 nm pulsed laser excitation with a fluence of 5 µJcm$^{-2}$ was used.

Figure 3(a) shows the evolution of the circular dichroism (CD) spectrum upon annealing of a pristine film. A large CD signal can be observed after annealing indicating the breaking of centrosymmetry in the crystallisation process. When spatially resolved, using Muller matrix imaging (MMPi) shown in Figure S.22, we see that this large CD signal is present across the spherulitic domains of the annealed thin-film and does not contain linear artefacts. As shown in Figure 1(d), the absorption edge and PL spectrum is red-shifted by 0.65 eV after annealing, which electronically resembles the S-TAT crystal, indicating that a similar electronic structure and chiral π-stacked packing is achieved by annealing the thin-films.

The CP-PL spectrum of the annealed films was measured using a home-built set-up utilising a rotating quarter-waveplate and fixed linear polariser(13, 25–28) (see Methods). We obtain a large dissymmetry factor, $g_{lum}$ ~ 10$^{-1}$, for both thin-films in the 500-650 nm domain. The transient response of the PL was probed using transient photoluminescence (TRPL) spectroscopy, shown in Figure 3(c)-3(d). The emission spectrum in annealed *S*-TAT(10wt%):DCzDCN thin-films shows little spectral changes over time, shown in Figure 3(c), it remains centered around 530 nm from 0 ns till 100 µs. Kinetic traces of pristine and annealed *S*-TAT(10wt%):DCzDCN thin-films, shown in Figure 3(d), shows the increase in PLQE and lifetime with annealing. The pristine films with blue emission have a PLQE of ~12% and lifetime of 2.7 ns, while the annealed films with green PL show a delayed lifetime of 23 ns with ~20% of PL delayed beyond 100 ns. We study the kinetics as a function of temperature, shown from 298K-10K in Figure 3(e), here we note that there is no change in the lifetime of prompt emission (t ~ 23 ns throughout) but interestingly the fraction of prompt PL increases at lower temperatures from ~75% at 298K to 87% at 10 K. A detailed analysis of the temperature dependent TRPL spectra is shown in Figures S.6-7.

**Circularly Polarised Organic Light-Emitting Diodes.**
We used successive vacuum-sublimation of charge injection, transport and recombination layers to make LED structures using 20 nm thick, 10wt% S-TAT doped thin-films as the recombination/emissive layer (EML). This architecture (23) is shown in Figure 4(a).

A broad range of different electron-transporting and hole-transporting host materials and neat *S*-TAT as EMLs (Figures S14-19, Table S2) was screened. In the case of a pure *S*-TAT EML, similar performances could be obtained compared to *S*-TAT(10wt%):CBP blends. Irrespective of the host-material, all devices exhibited green EL with EQE values beyond 10%, high current efficiencies and moderate to excellent brightness (Table S2). For both *S*-TAT(10%):CBP and *S*-TAT(10%):DCzDCN emitting layers (EMLs) dissymmetric EL was recorded, shown in Figure 4(c)-(d), with EL dissymmetry $g_{EL}$ ~ $10^{-1}$ (angular resolution of EL dissymmetry is shown in Figure S10). The best performing OLED was obtained from the ambipolar host DCzDCN. As shown in Figures 4(d)-(f), an $EQE_{max}$ of 15.7% was obtained with a $CE_{max}$ of 45.0 cd A$^{-1}$ at a low turn-on voltage of 2.2 V. Shown in Figure 4(e), the device exhibited bright electroluminescence with a $L_{max}$ as high as 57000 cd m$^{-2}$. The efficiency roll-off, shown in Figure 4(d), is significantly improved in DCzDCN containing devices compared to the CBP containing devices.

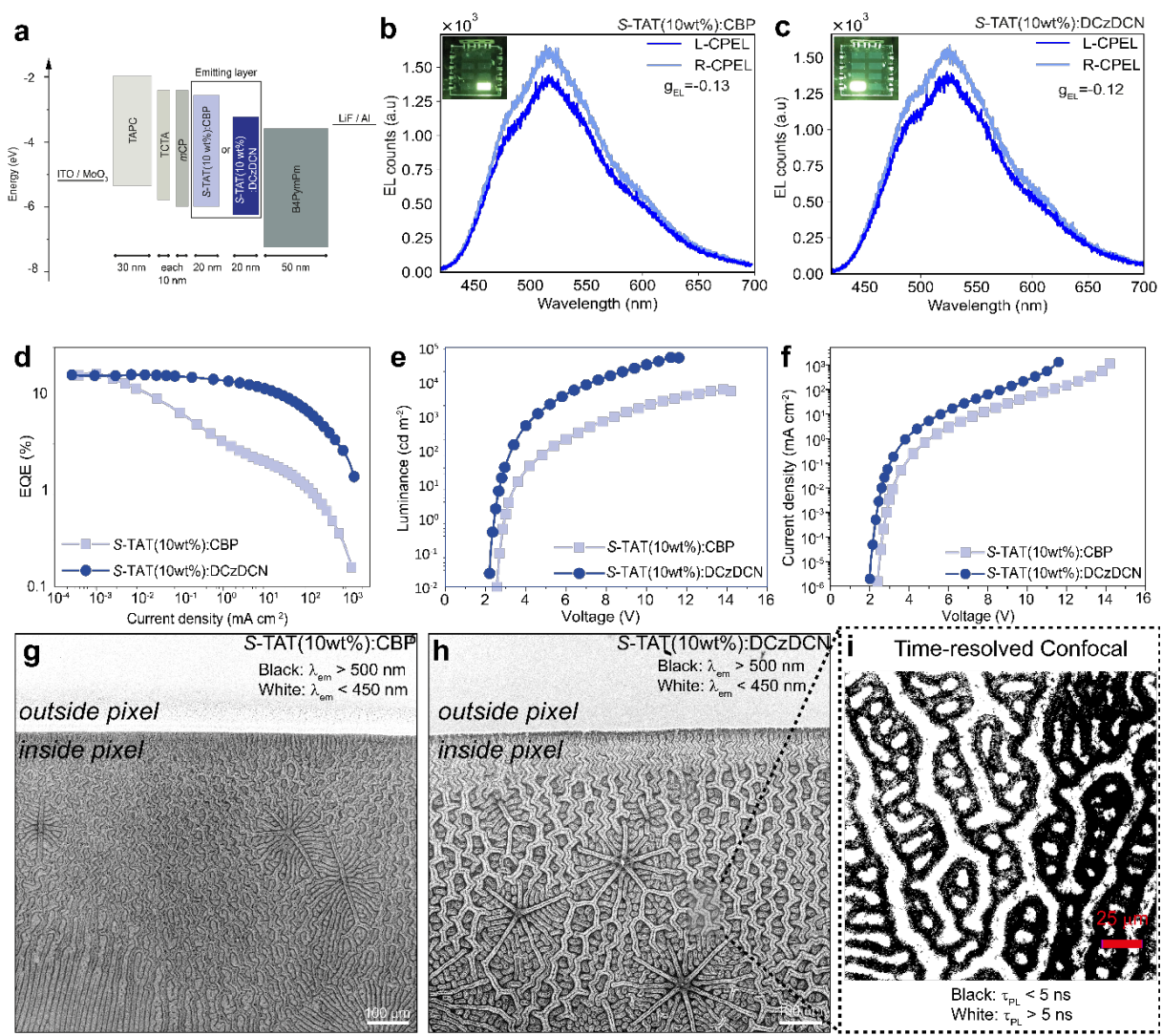

**Figure 4: Circularly polarised organic light emitting diodes (CPOLEDs). (a)** Device architecture used for the fabrication of CP-OLEDs using vacuum sublimation. **(b, c)** Circular polarised electroluminescence spectra for *S*-TAT(10%):CBP and *S*-TAT(10%):DCzDCN emitting layers. **(d)** External quantum efficiency, EQE, vs. current density curves for devices using *S*-TAT(10%):CBP (light-blue) and *S*-TAT(10%):DCzDCN (dark-blue) emitting layers. **(e)** Luminance vs. voltage curves for devices using *S*-TAT(10%):CBP (light-blue) and *S*-TAT(10%):DCzDCN (dark-blue) emitting layers. **(f)** Current-density vs. voltage curves for devices using *S*-TAT(10%):CBP (light-blue) and *S*-TAT(10%):DCzDCN (dark-blue) emitting layers. Steady state confocal images of the emitting layer in devices using **(g)** *S*-TAT(10%):CBP and **(h)** *S*-TAT(10%):DCzDCN as the emitting layer. A 450 nm excitation was used to capture emission image with $\lambda_{em}$ > 500 nm, a 380 nm excitation was used to capture emission image with $\lambda_{em}$ < 450 nm. **(i)** Time-resolved confocal image on a smaller region in the emitting layer where the regions with $\tau_{PL}$ > 5ns and $\tau_{PL}$ < 5ns are mapped. Images (g)-(h) shows that crystalline green domains are phase-segregated after being formed, these same domains show slow PL kinetics seen in (i).

We note that these LEDs can show an EQE of > 15%. Considering the PLQE (of up to 60% for the EML), the 3:1 spin-statistics of triplet:singlet excitons (29) and ideal light-coupling efficiency (30) we would expect only a maximum EQE of ~5% from our thin-films. We remarkably find this enhanced EQE for both *S*-TAT and *n*-TAT, and emphasise that both show local chiral domains, with equal amounts of left-handed and right-handed domains for *n*-TAT, as seen in Figure S.27. In contrast, non-assembled blue-emitting *phen*-TAT shows EQE a factor of about 4*x* lower than the PLQE, see Table S.2. We propose two possible explanations for this unexpected behaviour, both dependent on the presence of the supramolecular π-π packing:

*(i)* If both singlet and triplet excitons are generated by electron hole recombination, efficient luminescence would require a mechanism similar to that of TADF (thermally activated delayed fluorescence) OLEDs (31). This requires that the exchange energy between singlet and triplet be thermally accessible and is achieved in TADF systems by forming charge-transfer, CT, excitons. However, for isolated TAT molecules this cannot be the case. We therefore consider that the very large red shift of the singlet from isolated molecule to supramolecular stack (0.65 eV) brings the singlet close in energy to the triplet. The $T_1$ energy of TAT was recently reported (32) to be ~2.3 eV, close to the observed luminescence. There is evidence for some delayed emission following photoexcitation, up to 20% of the total, as presented in Figure 3(e). This slow emission may arise from thermal activation from the non-emissive triplet, as observed in TADF systems, though the component of slow emission seen here is smaller than for most TADF systems. We might expect that under pressure the singlet exciton would fall lower in energy with respect to the triplet exciton, but note the weak pressure dependence of emission energy and reduced PL yield under pressure, see Figures S.22-23.

*(ii)* Chiral-induced spin selectivity, CISS, has been identified as a mechanism for spin polarised injection and transport in chiral semiconductor structures(33–37). If this is active for electron and hole injection into the chiral TAT nanostructures present in the emissive layer, then the injected electron and hole would create only singlet excitons only.

Finally, we have investigated the microstructural origin of the emission regions within the TAT/host emissive layers. Confocal microscopy images on the EML of the driven pixels in the CP-OLEDs with *S*-TAT(10%):CBP and *S*-TAT(10%):DCzDCN, are shown in Figure 4(g)-(h). Upon 450 nm excitation, the regions with $\lambda_{PL}$ > 500 nm displays a large patterned network being formed which resembles the aftermath of viscoelastic phase-separated mixtures(*38*). Time-resolved confocal microscopy maps (tr-confocal), shown in Figure 4(i), shows us that the same pattern emerges with regions that have a lifetime, $\tau_{PL}$ > 5 ns. Conversely, when using a 350 nm excitation, regions with blue PL which are also the regions with $\tau_{PL}$ < 5 ns, is found in-between the green emitting regions. Comparing the microscopy results in Figure 4(g)-(h) to the PL spectrum in Figure 1(d) we can attribute the green emitting regions in the EML to the chiral helical, π-stacked *S*-TAT molecules which emit in green. Additionally, as seen in Figure 3(c)-(d) these green emitting domains have a lifetime of $\tau_{PL}$ ~ 23 ns which is verified spatially using the tr-confocal in Figure 4(i) (See Figures S.20-21 for more hosts and controls). When inspected closely, the emergent patterns formed by the $\lambda_{PL}$ > 500 nm regions resemble those obtained as an aftermath of viscoelastic phase separation of kinetically asymmetric mixtures (39*,* 40). In the thin-film, local heating can very quickly exceed 100°C during device operation, exceeding above the melting temperature of *S*-TAT but remaining below the melting temperature of the host. Under such conditions the quickly melting *S*-TAT phase can possibly reorganise and flow inside the host-matrix which triggers a strong bias for driving phase separation between the slow and fast components (41). In the case of *S*-TAT, the long-alkyl chains prevent host-guest interactions, providing a strong driving force for phase segregation and crystallisation of *S*-TAT into the preferred helical π-π stacked supramolecular assemblies (42, 43). When the interaction between the host and dopant is larger, such as when 10wt% *phen*-TAT is used (see Figure S.32), this phase separation is hampered resulting in only a slight red-shift in the PL and no emergent CP-PL from annealed films. The vacuum sublimed devices with *phen*-TAT(10wt%):CBP EML also have a much lower EQE ~ 3% which is close to the expected theroretical value, while obviously no CP-EL from the vacuum-sublimed OLEDs is observed. Furthermore, these *phen*-TAT doped OLEDs do not show any emergent phase-separated structures when subject to confocal microscopy. Finally, when investigating the n-TAT devices with *n*-TAT(10wt%):CBP, most optical properties and device characteristics, are similar to the S-TAT-based devices, except that a dissymmetry of $g_{EL}$ ~ 0 is obtained, in line with the racemic nature of the *n*-TAT stacks. These observations indicate that the nature of the N-substituent is directly involved in the formation of supramolecular structures in the films which is further linked to the photophysical and electrical properties of the thin-films.

A clear picture for the photophysical properties that we observe, as illustrated in Figure 1(c) emerged: *in-situ* crystallisation of *S*-TAT rich domains into the chiral supramolecular architecture is triggered by the local heating produced by electric driving. Crystallisation is propelled by the large entropy of the molten *S*-TAT phase (it has the lower $T_m$ in all the cases we have studied) which does not strongly interact with the slow-melting host matrix. The crystallisation leads to a *viscoelastic micro-phase segregation*. The extension of the singlet exciton across multiple *S*-TAT molecules lowers the singlet-triplet gap which red-shifts the emission and harvests the localised triplet states through fast *r*ISC over a small barrier. As discussed earlier, the strong chirality of the *S*-TAT supramolecular stacking imposes angular momentum selection rules on emitted photons laid out in Figure 2. This chirality can influence the spin-statistics of the electrically pumped excitons, by virtue of the CISS effect, by favouring the formation of singlet electron-hole pairs which can boost the luminescence efficiency.

In summary, we have set out a new scheme for the fabrication of efficient OLEDs which contain chiral structures that give circularly polarised emission without lowering the efficiency and brightness. Vacuum co-sublimation of guest-host systems selected to allow *in-situ* crystallisation to the chiral structure allows controllable thin film processing that is fully compatible with OLED fabrication. We consider this result has broad applications: circular polarised emission is of potential value for light control in LED displays, and this offers opportunities to explore spin selective transport processes(36, 37).

**Acknowledgements**

**R.H.F, H.H.C** and **R.C** received funding from the European Research Council under the European Union's Horizon 2020 research and innovation programme (Grant Agreement No. SCORS – 101020167). **E.W.M** received funding from the European Research Council under the European Union's Horizon 2020 research and innovation programme (Grant Agreement No. SYNMAT – 788618). **R.C**, **M.D.P** and **A.T** were supported by the European Union's Horizon 2020 project for funding under its research and innovation programme through Marie Skłodowska-Curie Actions (Grant Agreement No. 859752, HEL4CHIROLED). We acknowledge the Diamond Light Source B23 beamline for the MMP mapping. **E.W.M, S.C.J.M, J.v.T**, **B.W.L.v.d.B** and **G.V** were supported by the Dutch Ministry of Education, Culture and Science (Gravity program 024.001.035). **J.J.P.T** and **B.M** are supported by a EPSRC Programme Grant [EP/W017091/1]. **B.M** is also supported by a UKRI Future Leaders Fellowship [MR/V023926/1], by the Gianna Angelopoulos Programme for Science, Technology, and Innovation, and by the Winton Programme for the Physics of Sustainability. **S.S** was supported by the Bill & Melinda Gates Foundation [OPP1144]. **P.G** thanks the Cambridge Trust and the George and Lilian Schiff Foundation for a PhD scholarship and St John's College, Cambridge, for additional support. **Y.B** thanks the Winton Programme for Physics of Sustainability for funding. **S.T.K.** gratefully acknowledges funding from the European Union's Horizon 2020 research and innovation programme under the Marie Skłodowska-Curie grant agreement No. 101022365. **A.R** received funding from the European Research Council under the European Union's Horizon 2020 research and innovation programme (Grant Agreement No. 758826). **X.W.C.** thanks the Agency of Science, Technology and Research (A*STAR, Singapore) for funding from the National Science Scholarship.

**Author contributions**

**R.H.F** and **E.W.M** conceived the project. **R.C.** and **M.D.P** developed the project. **R.C** planned and conducted all the steady-state spectroscopy, transient spectroscopy and magneto-optical experiments and subsequent data analysis and interpretation. **M.D.P** designed and synthesized the molecules, assessed their purity and carried out the supramolecular assembly studies and subsequent data analysis and interpretation. **H.H.C** designed and fabricated the CP-OLEDs and all vacuum sublimed films for spectroscopy and performed steady-state photophysical measurements of the deposited films during fabrication and performed all device performance characterization with magneto and transient electroluminescent measurements. **T.B** built the circularly polarised optics, automated the process and helped in data acquisition. **J.J.P.T** and **K.W.C** performed first principles calculations on the crystal structures supervised by **B.M**. **S.S** conducted steady state confocal imaging. **P.G.** performed Raman spectroscopy and microscopy on the crystals and films. **Y.B** conducted photothermal-deflection spectroscopy on the crystals and thin-films. **E.G.** fabricated and measured CP-OLED devices. **X.W.C** performed time-resolved confocal microscopy on the thin-films. **J.v.T** performed AFM measurements. **B.W.L.v.d.B** performed SAXS measurements. **A.T** performed Muller-Matrix imaging experiments. **N.D** contributed cyclic voltammetry measurements. **D.M.D** performed pressure-dependent PL studies on thin-films and crystals supervised by **B.E**. **S.C.J.M** critically analysed, interpreted and designed the polarised optical studies. **G.V** assembly studies of the molecules. **P.G**, **Y.B** and **X.W.C** were supervised by **A.R**. **M.D.P** was supervised by **G.V** and **E.W.M**. **R.C**, **H.H.C and E.G.** were supervised by **R.H.F**. **R.H.F**, **E.W.M**, **R.C** and **M.D.P** wrote the manuscript with contributions from all authors.

**References**

1. K. Ariga, T. Mori, T. Kitao, T. Uemura, Supramolecular Chiral Nanoarchitectonics. *Advanced Materials* **32** (2020).

2. M. B. Baker, L. Albertazzi, I. K. Voets, C. M. A. Leenders, A. R. A. Palmans, G. M. Pavan, E. W. Meijer, Consequences of chirality on the dynamics of a water-soluble supramolecular polymer. *Nat Commun* **6**, 6234 (2015).

3. P. Besenius, G. Portale, P. H. H. Bomans, H. M. Janssen, A. R. A. Palmans, E. W. Meijer, Controlling the growth and shape of chiral supramolecular polymers in water. *Proceedings of the National Academy of Sciences* **107**, 17888–17893 (2010).

4. S. M. C. Schoenmakers, A. J. H. Spiering, S. Herziger, C. Böttcher, R. Haag, A. R. A. Palmans, E. W. Meijer, Structure and Dynamics of Supramolecular Polymers: Wait and See. *ACS Macro Lett* **11**, 711–715 (2022).

5. T. F. A. De Greef, M. M. J. Smulders, M. Wolffs, A. P. H. J. Schenning, R. P. Sijbesma, E. W. Meijer, Supramolecular Polymerization. *Chem Rev* **109**, 5687–5754 (2009).

6. J. H. K. K. Hirschberg, L. Brunsveld, A. Ramzi, J. A. J. M. Vekemans, R. P. Sijbesma, E. W. Meijer, Helical self-assembled polymers from cooperative stacking of hydrogen-bonded pairs. *Nature* **407**, 167–170 (2000).

7. C. Kulkarni, A. K. Mondal, T. K. Das, G. Grinbom, F. Tassinari, M. F. J. Mabesoone, E. W. Meijer, R. Naaman, Highly Efficient and Tunable Filtering of Electrons' Spin by Supramolecular Chirality of Nanofiber-Based Materials. *Advanced Materials* **32** (2020).

8. M. Zhang, Q. Guo, Z. Li, Y. Zhou, S. Zhao, Z. Tong, Y. Wang, G. Li, S. Jin, M. Zhu, T. Zhuang, S.-H. Yu, Processable circularly polarized luminescence material enables flexible stereoscopic 3D imaging. *Sci Adv* **9** (2023).

9. N. Nishizawa, K. Nishibayashi, H. Munekata, Pure circular polarization electroluminescence at room temperature with spin-polarized light-emitting diodes. *Proceedings of the National Academy of Sciences* **114**, 1783–1788 (2017).

10. P. Stachelek, L. MacKenzie, D. Parker, R. Pal, Circularly polarised luminescence laser scanning confocal microscopy to study live cell chiral molecular interactions. *Nat Commun* **13**, 553 (2022).

11. B. M. W. Langeveld-Voss, R. A. J. Janssen, M. P. T. Christiaans, S. C. J. Meskers, H. P. J. M. Dekkers, E. W. Meijer, Circular Dichroism and Circular Polarization of Photoluminescence of Highly Ordered Poly{3,4-di[( *S* )-2-methylbutoxy]thiophene}. *J Am Chem Soc* **118**, 4908–4909 (1996).

12. E. Peeters, M. P. T. Christiaans, R. A. J. Janssen, H. F. M. Schoo, H. P. J. M. Dekkers, E. W. Meijer, Circularly Polarized Electroluminescence from a Polymer Light-Emitting Diode. *J Am Chem Soc* **119**, 9909–9910 (1997).

13. Y.-H. Kim, Y. Zhai, H. Lu, X. Pan, C. Xiao, E. A. Gaulding, S. P. Harvey, J. J. Berry, Z. V. Vardeny, J. M. Luther, M. C. Beard, Chiral-induced spin selectivity enables a room-temperature spin light-emitting diode. *Science (1979)* **371**, 1129–1133 (2021).

14. D. Di Nuzzo, C. Kulkarni, B. Zhao, E. Smolinsky, F. Tassinari, S. C. J. Meskers, R. Naaman, E. W. Meijer, R. H. Friend, High Circular Polarization of Electroluminescence Achieved *via* Self-Assembly of a Light-Emitting Chiral Conjugated Polymer into Multidomain Cholesteric Films. *ACS Nano* **11**, 12713–12722 (2017).

15. Y. Yang, B. Rice, X. Shi, J. R. Brandt, R. Correa da Costa, G. J. Hedley, D.-M. Smilgies, J. M. Frost, Ifor. D. W. Samuel, A. Otero-de-la-Roza, E. R. Johnson, K. E. Jelfs, J. Nelson, A. J. Campbell, M. J. Fuchter, Emergent Properties of an Organic Semiconductor Driven by its Molecular Chirality. *ACS Nano* **11**, 8329–8338 (2017).

16. J. R. Brandt, X. Wang, Y. Yang, A. J. Campbell, M. J. Fuchter, Circularly Polarized Phosphorescent Electroluminescence with a High Dissymmetry Factor from PHOLEDs Based on a Platinahelicene. *J Am Chem Soc* **138**, 9743–9746 (2016).

17. F. Zinna, U. Giovanella, L. Di Bari, Highly Circularly Polarized Electroluminescence from a Chiral Europium Complex. *Advanced Materials* **27**, 1791–1795 (2015).

18. J. P. Schmidtke, J.-S. Kim, J. Gierschner, C. Silva, R. H. Friend, Optical Spectroscopy of a Polyfluorene Copolymer at High Pressure: Intra- and Intermolecular Interactions. *Phys Rev Lett* **99**, 167401 (2007).

19. J. R. Brandt, X. Wang, Y. Yang, A. J. Campbell, M. J. Fuchter, Circularly Polarized Phosphorescent Electroluminescence with a High Dissymmetry Factor from PHOLEDs Based on a Platinahelicene. *J Am Chem Soc* **138**, 9743–9746 (2016).

20. M. Kira, S. W. Koch, Many-body correlations and excitonic effects in semiconductor spectroscopy. *Prog Quantum Electron* **30**, 155–296 (2006).

21. J. J. P. Thompson, D. Muth, S. Anhäuser, D. Bischof, M. Gerhard, G. Witte, E. Malic, Singlet-exciton optics and phonon-mediated dynamics in oligoacene semiconductor crystals. *Natural Sciences* **3** (2023).

22. Q. Zhang, B. Li, S. Huang, H. Nomura, H. Tanaka, C. Adachi, Efficient blue organic light-emitting diodes employing thermally activated delayed fluorescence. *Nat Photonics* **8**, 326–332 (2014).

23. H. Cho, S. Gorgon, H. Hung, J. Huang, Y. Wu, F. Li, N. C. Greenham, E. W. Evans, R. H. Friend, Efficient and Bright Organic Radical Light-Emitting Diodes with Low Efficiency Roll-Off. *Advanced Materials* **35** (2023).

24. Y. J. Cho, K. S. Yook, J. Y. Lee, A Universal Host Material for High External Quantum Efficiency Close to 25% and Long Lifetime in Green Fluorescent and Phosphorescent OLEDs. *Advanced Materials* **26**, 4050–4055 (2014).

25. B. Baguenard, A. Bensalah-Ledoux, L. Guy, F. Riobé, O. Maury, S. Guy, Theoretical and experimental analysis of circularly polarized luminescence spectrophotometers for artifact-free measurements using a single CCD camera. *Nat Commun* **14**, 1065 (2023).

26. Y.-H. Kim, Y. Zhai, E. A. Gaulding, S. N. Habisreutinger, T. Moot, B. A. Rosales, H. Lu, A. Hazarika, R. Brunecky, L. M. Wheeler, J. J. Berry, M. C. Beard, J. M. Luther, Strategies to Achieve High Circularly Polarized Luminescence from Colloidal Organic–Inorganic Hybrid Perovskite Nanocrystals. *ACS Nano* **14**, 8816–8825 (2020).

27. F. Zinna, T. Bruhn, C. A. Guido, J. Ahrens, M. Bröring, L. Di Bari, G. Pescitelli, Circularly Polarized Luminescence from Axially Chiral BODIPY DYEmers: An Experimental and Computational Study. *Chemistry – A European Journal* **22**, 16089–16098 (2016).

28. M. Morgenroth, M. Scholz, M. J. Cho, D. H. Choi, K. Oum, T. Lenzer, Mapping the broadband circular dichroism of copolymer films with supramolecular chirality in time and space. *Nat Commun* **13**, 210 (2022).

29. C. J. Bardeen, The Structure and Dynamics of Molecular Excitons. *Annu Rev Phys Chem* **65**, 127–148 (2014).

30. R. Meerheim, M. Furno, S. Hofmann, B. Lüssem, K. Leo, Quantification of energy loss mechanisms in organic light-emitting diodes. *Appl Phys Lett* **97** (2010).

31. H. Uoyama, K. Goushi, K. Shizu, H. Nomura, C. Adachi, Highly efficient organic light-emitting diodes from delayed fluorescence. *Nature* **492**, 234–238 (2012).

32. C.-Y. Lin, C.-H. Hsu, C.-M. Hung, C.-C. Wu, Y.-H. Liu, E. H.-C. Shi, T.-H. Lin, Y.-C. Hu, W.-Y. Hung, K.-T. Wong, P.-T. Chou, Entropy-driven charge-transfer complexation yields thermally activated delayed fluorescence and highly efficient OLEDs. *Nat Chem* **16**, 98–106 (2024).

33. B. Göhler, V. Hamelbeck, T. Z. Markus, M. Kettner, G. F. Hanne, Z. Vager, R. Naaman, H. Zacharias, Spin Selectivity in Electron Transmission Through Self-Assembled Monolayers of Double-Stranded DNA. *Science (1979)* **331**, 894–897 (2011).

34. R. Naaman, Y. Paltiel, D. H. Waldeck, Chiral molecules and the electron spin. *Nat Rev Chem* **3**, 250–260 (2019).

35. A. Chiesa, A. Privitera, E. Macaluso, M. Mannini, R. Bittl, R. Naaman, M. R. Wasielewski, R. Sessoli, S. Carretta, Chirality-Induced Spin Selectivity: An Enabling Technology for Quantum Applications. *Advanced Materials* **35** (2023).

36. B. P. Bloom, Y. Paltiel, R. Naaman, D. H. Waldeck, Chiral Induced Spin Selectivity. *Chem Rev* **124**, 1950–1991 (2024).

37. B. Göhler, V. Hamelbeck, T. Z. Markus, M. Kettner, G. F. Hanne, Z. Vager, R. Naaman, H. Zacharias, Spin Selectivity in Electron Transmission Through Self-Assembled Monolayers of Double-Stranded DNA. *Science (1979)* **331**, 894–897 (2011).

38. H. Tanaka, Viscoelastic phase separation. *Journal of Physics: Condensed Matter* **12**, R207–R264 (2000).

39. H. Tanaka, T. Araki, T. Koyama, Y. Nishikawa, Universality of viscoelastic phase separation in soft matter. *Journal of Physics: Condensed Matter* **17**, S3195–S3204 (2005).

40. H. Tanaka, T. Araki, Viscoelastic phase separation in soft matter: Numerical-simulation study on its physical mechanism. *Chem Eng Sci* **61**, 2108–2141 (2006).

41. A. Brunk, B. Dünweg, H. Egger, O. Habrich, M. Lukáčová-Medvid'ová, D. Spiller, Analysis of a viscoelastic phase separation model. *Journal of Physics: Condensed Matter* **33**, 234002 (2021).

42. W. M. Jacobs, D. W. Oxtoby, D. Frenkel, Phase separation in solutions with specific and nonspecific interactions. *J Chem Phys* **140** (2014).

43. D. Frenkel, A. A. Louis, Phase separation in binary hard-core mixtures: An exact result. *Phys Rev Lett* **68**, 3363–3365 (1992).